\documentclass[pra,
                twocolumn,
                preprintnumbers,
                amsmath,
                amssymb,
                showpacs,
                superscriptaddress]{revtex4-1}
\usepackage{graphicx} 
\usepackage{dcolumn}  
\usepackage{bm}       
\usepackage{amsfonts}
\usepackage{mathtools}
\usepackage[colorlinks=true,linkcolor=blue,citecolor=red,urlcolor=blue]{hyperref}
\usepackage{color}
\usepackage{natbib}
\newcommand{\vscf}{\mbox{$\delta V$}}
\newcommand{\eq}[1]{Eq.~(\ref{#1})}

\begin{document}

\title{Effects of exchange-correlation potentials on the density functional description of C$_{60}$ versus C$_{240}$ photoionization}

\author{Jinwoo Choi}
\affiliation{%
Department of Natural Sciences, D.L.\ Hubbard Center for Innovation and Entrepreneurship,
Northwest Missouri State University, Maryville, Missouri 64468, USA}

\author{EonHo Chang}
\affiliation{%
Department of Natural Sciences, D.L.\ Hubbard Center for Innovation and Entrepreneurship,
Northwest Missouri State University, Maryville, Missouri 64468, USA}

\author{Dylan M. Anstine}
\affiliation{%
Department of Natural Sciences, D.L.\ Hubbard Center for Innovation and Entrepreneurship,
Northwest Missouri State University, Maryville, Missouri 64468, USA}

\author{Mohamed El-Amine Madjet}
\affiliation{%
Qatar Environment and Energy Research Institute, Hamad Bin Khalifa University, Qatar Foundation, P.O Box 5825, 
Doha, Qatar}

\author{Himadri S. Chakraborty}
\email[]{himadri@nwmissouri.edu}
\affiliation{%
Department of Natural Sciences, D.L.\ Hubbard Center for Innovation and Entrepreneurship,
Northwest Missouri State University, Maryville, Missouri 64468, USA}

\date{\today}



\begin{abstract}
We study the photoionization properties of the C$_{60}$ versus C$_{240}$ molecule in a spherical jellium frame of density functional method. Two different approximations to the exchange-correlation (xc) functional are used: (i) The Gunnerson-Lundqvist parametrization [Phys.\ Rev.\ B {\bf 13}, 4274 (1976)] with an explicit  correction for the electron self-interaction (SIC) and (ii) a gradient-dependent augmentation of (i) by using the van Leeuwen and Baerends model potential [Phys.\ Rev.\ A {\bf 49}, 2421 (1994)], in lieu of SIC, to implicitly restore electrons' asymptotic properties. Ground state results from the two schemes for both molecules show differences in the shapes of mean-field potentials and bound-level properties. The choice of a xc scheme also significantly alters the dipole single-photoionization cross sections obtained by an {\em ab\,initio} method that incorporates linear-response dynamical correlations. Differences in the structures and ionization responses between C$_{60}$ and C$_{240}$ uncover the effect of molecular size on the underlying physics. Analysis indicates that the collective plasmon resonances with the gradient-based xc-option produce results noticeably closer to the experimental data available for C$_{60}$.
\end{abstract}

\maketitle 

\section{Introduction}
Fullerene molecules are a highly stable form of nanoscopic carbon allotrope that can exist at room temperature. Therefore, they are routinely attractive candidates for spectroscopic studies in understanding aspects of fundamental physics both in their vapor and condensed matter phases. Technologically also, fullerenes hold the prospect of exciting applications in solid state quantum computations~\cite{harneit2007,ju2011}, improving the superconducting ability of materials~\cite{takeda2006}, biomedical fields~\cite{melanko2009}, contrast-enhancement research for magnetic resonance imaging (MRI), and improving organic photovoltaic devices~\cite{ross2009}. Therefore, investigations of the response of these materials to radiations are valuable. One direction of these studies is to understand the collective response of fullerene electrons to relatively low-energy photons. In an infinite system like graphite, the incoming oscillatory electric field induces plane-wave type plasma oscillations in the electron cloud within the system's translational symmetry. This can only quantize a surface plasmon quasi-particle, but not the longitudinal (compressional) volume plasmon, since light is a transverse wave. But when the medium has a boundary, the broken translational symmetry enables the plasma wave to reflect and induce other eigen modes of oscillations, including the volume quantization. In particular, for finite systems with boundaries in all directions, such as fullerenes and metallic nanoclusters, photospectroscopy reveals multiple plasmons that were measured~\cite{scully05,xia2009}. The photoelectron angular distribution asymmetry~\cite{maurat2009} and the emission time delay~\cite{barillot2015} at the surface plasmon of C$_{60}$ also predicted interesting behaviors. The other direction of fullerene studies involves the response to photons whose energy is higher than the plasmon excitation energies. These photons with their shorter wavelengths begin to resolve the fullerene molecular geometry, entering the spectral region of photoelectron diffraction. This effect results into the occurrence of a series of cavity minima observed in the ionization spectra as the integer multiples of the photoelectron half wavelength fit the molecular redii at certain energies~\cite{ruedel02}. The effect also accompanies a beating modulation in the ionization spectra as a signature of C$_{60}$ molecular width~\cite{korica05}. Emission delay spectroscopy predicted structures at these minima~\cite{magrakvelidze2015cavitymin}.

Since the first observation of C$_{60}$ giant plasmon resonance~\cite{hertel92}, theoretical studies with various levels of approximation and success formed a large body of published research, an account of which up until 2008 can be found in Ref.\,[\onlinecite{madjet-jpb2008}]. After 2008, there have been mainly two lines of theoretical calculations that attempted to account for the atomistic details of the fullerene carbon-core on a truncated icosahedral geometry. One involves the geometric optimization of the C$_{60}$ structure by the commercially available DMol3 software followed by the calculation of Kohn-Sham ground state and then its linear response to the incoming radiation~\cite{chen2012}. The other uses the general access OCTOPUS software to directly solve the time-dependent density functional equations for excited C$_{60}$ to subsequently Fourier transform the density fluctuation to obtain the dynamical structure factor utilized to derive the electron energy-loss signal~\cite{schueler2015}. However, in spite of these important new developments, the jellium approximation to C$^{4+}$ ion-core, a model based on which we have developed a linear-response density functional methodology known as the time-dependent local density approximation (TDLDA), has seen a significant range of success over last several years and continues to remain relevant~\cite{chakraborty2015}. This is because of the ease and transparency of this model to capture the primary, robust observable effects and to access the key physics that underpins the photo-dynamics and related spectroscopy. Let us cite two sets of results from our methods that directly connected the experiments: (i) Our calculations have predicted the photoionization of a second plasmon at a higher energy whose first observation was reported in our joint publications~\cite{scully05,scully07} with the experimental group for gas phase C$_{60}$ anions; a subsequent experiment accessed this new plasmon even for the neutral C$_{60}$~\cite{rein04}. (ii) Another experiment-theory joint study of ours revealed oscillations in C$_{60}$ valence photoemissions providing trains of diffraction minima mentioned above~\cite{ruedel02}. Besides these pivotal results, our jellium-based study also extended to the photoionization of several atomic endofullerene molecules~\cite{madjet2007,chakraborty2008,madjet2010,maser2012,javani2014} and the C$_{60}$@C$_{240}$ buckyonion~\cite{mccune2011}. For some of these fullerene systems, TDLDA investigations of the photoemission time-delay~\cite{barillot2015,magrakvelidze2015cavitymin,dixit2013} and multitudes of resonant inter-Coulombic decay processes~\cite{javani2014rhaicd,magrakvelidze2016,de2016} were also carried out with reasonable success.

One limitation of the Kohn-Sham density functional method is its approximate treatment of the electron exchange. This is because the exchange interaction can only be fully treated in a non-local theory such as Hartree-Fock (HF) that exactly cancels out all self-interactions, restoring correct $1/r$ behavior at $r\rightarrow \infty$. In most of our previous calculations involving C$_{60}$ and its derivative endo-C$_{60}$ compounds we used a widely utilized approximate scheme~\cite{gunnerson76} of exchange-correlation (xc) functional augmented by an orbit-by-orbit elimination of self-interaction~\cite{madjet01} originally proposed by Perdew and Zunger~\cite{perdew1981}. A different scheme is to use the gradient-corrected xc potential of van Leeuwen and Baerends~\cite{van1994exchange} that intrinsically approximates the correct long distance properties. While we adopted the latter in some of our most recent works, no detailed study on the comparative abilities between the two schemes has yet been made. This is the primary objective of the current work that considers the photoionization of the fullerene molecule. Along with C$_{60}$, a larger spherical fullerene, C$_{240}$, has also been considered to further broaden the scope of the comparison. Significant differences from the choice of the xc treatment, both in ground and photoionization descriptions, are uncovered. Improved agreement of C$_{60}$ plasmonic spectrum from the gradient-corrected xc approach with the measured data is found. 
   
This paper is structured as follows. Section II includes three subsections: A) the description of jellium core ground state structures with brief accounts of two xc parametrization schemes, B) comparison of ground state numerical results between two schemes and between two fullerenes, and C) the essentials of the method that incorporates electron correlations in responding to the radiation;  Section III compares the results of the valence (subsection A) and total (subsection B) photoemissions, as well as a comparisons with available measurements for C$_{60}$ (subsection C). Conclusions are presented in Section IV. 

\section{Essentials of the Method}

\subsection{LDA exchange-correlation functionals}

The details of the method follow the framework as described in Ref.\ [\onlinecite{madjet-jpb2008}]. The jellium potentials, $V_{\mbox{\scriptsize jel}}(\mathbf{r})$, representing 60 and 240 C$^{4+}$ ions, respectively for C$_{60}$ and C$_{240}$, are constructed by smearing the total positive charge over spherical shells with radius $R$ and thickness $\Delta$. $R$ is taken to be the known radius of each molecule: 3.54 \AA\, for C$_{60}$ and 7.14 \AA\, for C$_{240}$. A constant pseudopotential $V_0$ is added to the jellium for quantitative accuracy~\cite{puska93}. The Kohn-Sham equations for systems of 240 and 960 electrons, made up of four valence ($2s^22p^2$) electrons from each carbon atom, are then solved to obtain the single electron ground state orbitals in the local density approximation (LDA). The parameters $V_0$ and $\Delta$ are determined by requiring both charge neutrality and obtaining the experimental value (for C$_{60}$) and the known theoretical value (for C$_{240}$) of the first ionization thresholds. The values of $\Delta$ and the binding energies of the highest occupied molecular orbital (HOMO) and HOMO-1 levels of both systems are given in Table\,1.

Using the single-particle density $\rho({\bf r})$ the LDA potential can be written as,
\begin{equation}\label{lda-pot}
V_{\scriptsize \mbox{LDA}}(\mathbf{r}) = V_{\mbox{\scriptsize jel}}(\mathbf{r}) + \int d\mathbf{r}'\frac{\rho(\mathbf{r}')}{|\mathbf{r}-\mathbf{r}'|} + V_{\scriptsize \mbox{XC}}[\rho(\mathbf{r})],
\end{equation}
where the 2nd and 3rd terms on the right are the direct and xc components. In one scheme, $V_{\scriptsize \mbox{XC}}$ is parametrized directly from $\rho(\mathbf{r})$ by the following formula~\cite{gunnerson76}:
\begin{eqnarray}\label{gl}
V_{\scriptsize \mbox{XC}}[\rho(\mathbf{r})] & = & -\left(\frac{3\rho(\mathbf{r})}{\pi}\right)^{1/3} \nonumber \\
 &-& 0.0333\log\left[1 + 11.4\left(\frac{4\pi\rho(\mathbf{r})}{3}\right)^{1/3}\right],
\end{eqnarray}
in which the first term on the right is exactly derivable by a variational approach from the HF exchange energy of a uniform electron system with a uniform positively charged background and the second term is the so called correlation potential, a quantity not borne in HF formalism. In addition, we include an appropriate correction to eliminate unphysical electron self-interactions for the $i$-th subshell that renders the LDA potential orbital-specific~\cite{madjet99,madjet01},
\begin{eqnarray}\label{sic}
V^i({\bf r}) & = & V_{\mbox{\scriptsize jel}}({\bf r}) + \int d{\bf r'} \frac{\rho({\bf r'}) 
               - \rho_i ({\bf r'})}{{\bf r}-{\bf r'}} \nonumber \\
               &+& (V_{\mbox{\scriptsize XC}}[\rho({\bf r})] - V_{\mbox{\scriptsize XC}}[\rho_i({\bf r})]).
\end{eqnarray}
This correction approximately captures the electron's long range properties. We use the acronym SIC to refer this model.

The other alternative account for xc-functional that utilizes \eq{gl} but further refines it by adding a parametrized potential~\cite{van1994exchange} in terms of the density and its gradient $\nabla \rho$ as follows,
\begin{equation}\label{lb94}
V_{\mbox{\scriptsize LB}} = -\beta{[\rho({\bf r})]^{1/3}}\frac{X^2}{1+3\beta X \sinh^{-1}(X)},
\end{equation}
where $\beta$ is adjustable and $X=[\nabla \rho]/\rho^{4/3}$. This scheme, termed as LB94, is known to have lead to a considerable improvement in the asymptotic behavior of the electron when compared to the exact Kohn-Sham potentials calculated from correlated densities. Consequently, this model is expected to also significantly improve the quality of both the excited and continuum spectra. 

\begin{figure*}
\includegraphics[width=13cm]{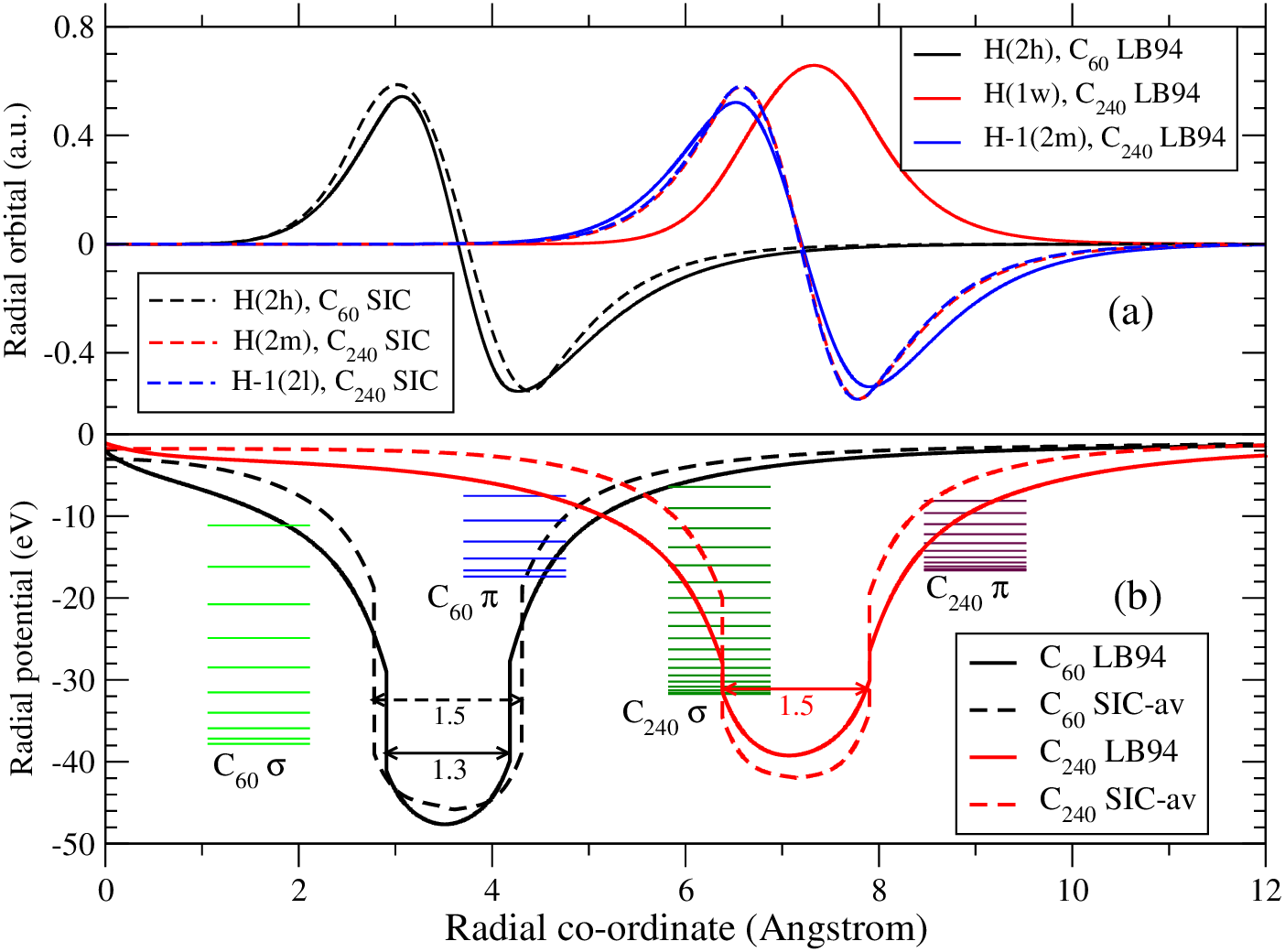}
\caption{(Color online) (a) Ground state radial wavefunctions for C$_{60}$ HOMO (H) levels and both C$_{240}$ HOMO and HOMO-1 levels calculated in SIC and LB94. (b) Corresponding radial potentials are shown. Shell widths are identified. Energy bands of $\sigma$ and $\pi$ characters (see text), obtained only in LB94, are illustrated} \label{fig1}
\end{figure*}

\subsection{Ground states of C$_{60}$ and C$_{240}$: SIC versus LB94}

We show the ground state radial potentials of C$_{60}$ and C$_{240}$ obtained via both SIC and LB94 in Fig.\,1(b) where the SIC curves, labeled as SIC-av, are occupancy-weighted average over all the subshells. This particular shape of the potentials earlier interpreted multiple frequencies in the Fourier transform of the measured photoelectron spectra of C$_{60}$~\cite{ruedel02}. Yet, note the differences in details from SIC to LB94: (i) For C$_{60}$, to retain the exact same configuration of occupied states optimized earlier~\cite{madjet-jpb2008} based on a number of experimental findings~\cite{vos97,weaver91}, the LB94 potential gets slightly narrower (see Table\,1) and deeper but with more widening of the wings on either side of the shell. (ii) While these general shapes also hold good for C$_{240}$, we note the following. In the absence of enough experimental information, C$_{240}$ ground states were optimized by requiring identical widths and similar first ionization energies (Table\,1) for both SIC and LB94. This alters some properties of occupied configuration that includes the LB94 HOMO to be of $\sigma$ character (a level with no radial node) $1w$ with a very high angular momentum $\ell=18$ as opposed to a $\pi$ level (with one node) $2m$ in SIC of a lower $\ell=10$ [see Fig.\,1(a)]. The direct repercussion of this change on their photoionization cross sections will be discussed below in section III A. Fig.\,1(a) illustrates the general differences, SIC versus LB94, of some valence radial wavefunctions. (iii) Finally, the potential depth decreases from C$_{60}$ to C$_{240}$ even though the latter accommodates four times more electrons than the former. Why does this happen? To answer, we need to bear in mind that the effective radial potential also includes the angular momentum dependent centrifugal barrier part $\ell(\ell+1)/2r^2$ which varies slower as a function of $r$ over the C$_{240}$ shell region that is radially farther from C$_{60}$, creating more ``energy-room" for larger C$_{240}$. Indeed, a far denser angular momentum manifold of $\pi$ and $\sigma$ energy bands are generally found for C$_{240}$ as seen for LB94 bands presented in Fig.\,1(b).
%
\begin{table}
\caption{Molecular shell-widths, and quantum characters ($n\ell$) in harmonic oscillator notations and binding energies (BE) of HOMO and HOMO-1 levels of C$_{60}$ and C$_{240}$. The values in parenthesis correspond to LB94 results only when different from SIC.}
\label{tab1}       
\begin{tabular}{c|ccccc}
\hline\noalign{\smallskip}
    & $\Delta$ (\AA) & HOMO & BE$_{\mbox{\scriptsize H}}$ (eV) &  HOMO-1  & BE$_{\mbox{\scriptsize H-1}}$ (eV)\\
\noalign{\smallskip}\hline\noalign{\smallskip}
C$_{60}$  & 1.50 (1.30)  &     $2h$    &      $-7.51$      &    $2g$     &      $-10.6$      \\
C$_{240}$ &     1.50     & $2m$ ($1w$) & $-6.47$ ($-6.43$) & $2l$ ($2m$) & $-7.98$ ($-8.12$) \\
\noalign{\smallskip}\hline
\end{tabular}
\end{table}

Our SIC and LB94 descriptions of C$_{60}$ also produced static dipole polarizability (SDP) values of respectively 92.8 \AA$^3$ and 114 \AA$^3$ which are reasonably close to the measured value of 76.5 $\pm$ 8 \AA$^3$~\cite{compagnon2001}, particularly given that the jellium model disregards the molecular core vibration. Likewise, our calculated values of SDP for C$_{240}$ are 565 \AA$^3$ and 638 \AA$^3$, respectively for SIC and LB94. The slight increase in SDP from SIC to LB94 for both fullerenes is due to a somewhat higher spill-out electron density in LB94. This spill-out can be recognized by noticing the LB94 potential in Fig.\,1 being a bit wider at the top causing slight outward spreads of the radial wavefunctions.

\subsection{TDLDA dynamical Response}

A time-dependent LDA (TDLDA) approach~\cite{madjet-jpb2008} is used to calculate the dynamical response of the compounds to the external dipole field $z$. In this method, the photoionization cross section corresponding to a bound-to-continuum dipole transition $n\ell\rightarrow k\ell^\prime$ is
\begin{equation}\label{cross-pi}
\sigma_{n\ell\rightarrow k\ell'} \sim |\langle k\ell'|z+\vscf|n\ell\rangle|^2,
\end{equation}
where the matrix element $M = D + \langle\vscf\rangle$, with $D$ being the independent-particle LDA matrix element; obviously, $D$ solely yields the LDA cross section. Here $\delta V$ represents the complex induced potential that accounts for electron correlations. In the TDLDA, $z + \delta V$ is proportional to the induced frequency-dependent changes in the electron density~\cite{madjet-jpb2008,magrakvelidze2016}.
This change is 
\begin{equation}\label{ind-dens}
\delta \rho (\mathbf{r}^{\prime}; \omega) = \int \chi (\mathbf{r}, \mathbf{r}^{\prime}; \omega)
z  d\mathbf{r},
\end{equation}
where the full susceptibility $\chi$ builds the dynamical correlation from the LDA susceptibilities, 
\begin{eqnarray}\label{suscep}
\chi^{0} (\mathbf{r},\mathbf{r}^{\prime };\omega) &=&\sum_{nl}^{occ}\phi _{nl}^{*}
(\mathbf{r})\phi _{nl}(\mathbf{r}^{\prime })\ G(\mathbf{r},\mathbf{r}^{\prime };\epsilon
_{nl}+\omega)  \nonumber \\
&+&\sum_{nl}^{occ}\phi _{nl}(\mathbf{r})\phi _{nl}^{*}(\mathbf{r}^{\prime })\ G^*
(\mathbf{r},\mathbf{r}^{\prime };\epsilon _{nl}-\omega)  
\end{eqnarray}
via   the matrix equation $\chi = \chi^0[1-(\partial V/\partial \rho)\chi^0]^{-1}$ involving the variation of the ground-state potential $V$ with respect to the ground-state density $\rho$. The radial components of the full Green's functions in \eq{suscep} are constructed with the regular ($f_L$) and irregular ($g_L$) solutions of the homogeneous radial equation 
\begin{equation}\label{radial-eq}
\left( \frac{1}{r^2} \frac{\partial}{\partial r} r^2 \frac{\partial}{\partial r} 
     - \frac{L(L+1)}{r^2} - V_{\mbox{\scriptsize{LDA}}} 
     + E \right) f_L(g_L) (r;E) = 0
\end{equation}
as
\begin{equation}\label{green}
G_{L}(r,r^{\prime };E)=\frac{2f_{L}(r_{<};E)h_{L}(r_{>};E)}{W [f_{L},h_{L}]}  
\end{equation}
where $W$ represents the Wronskian and $h_{L} = g_{L} + i\; f_{L}$. Obviously, TDLDA thus includes the dynamical correlation by improving upon the mean-field LDA basis.

\begin{figure}[h!]
\includegraphics[width=8cm]{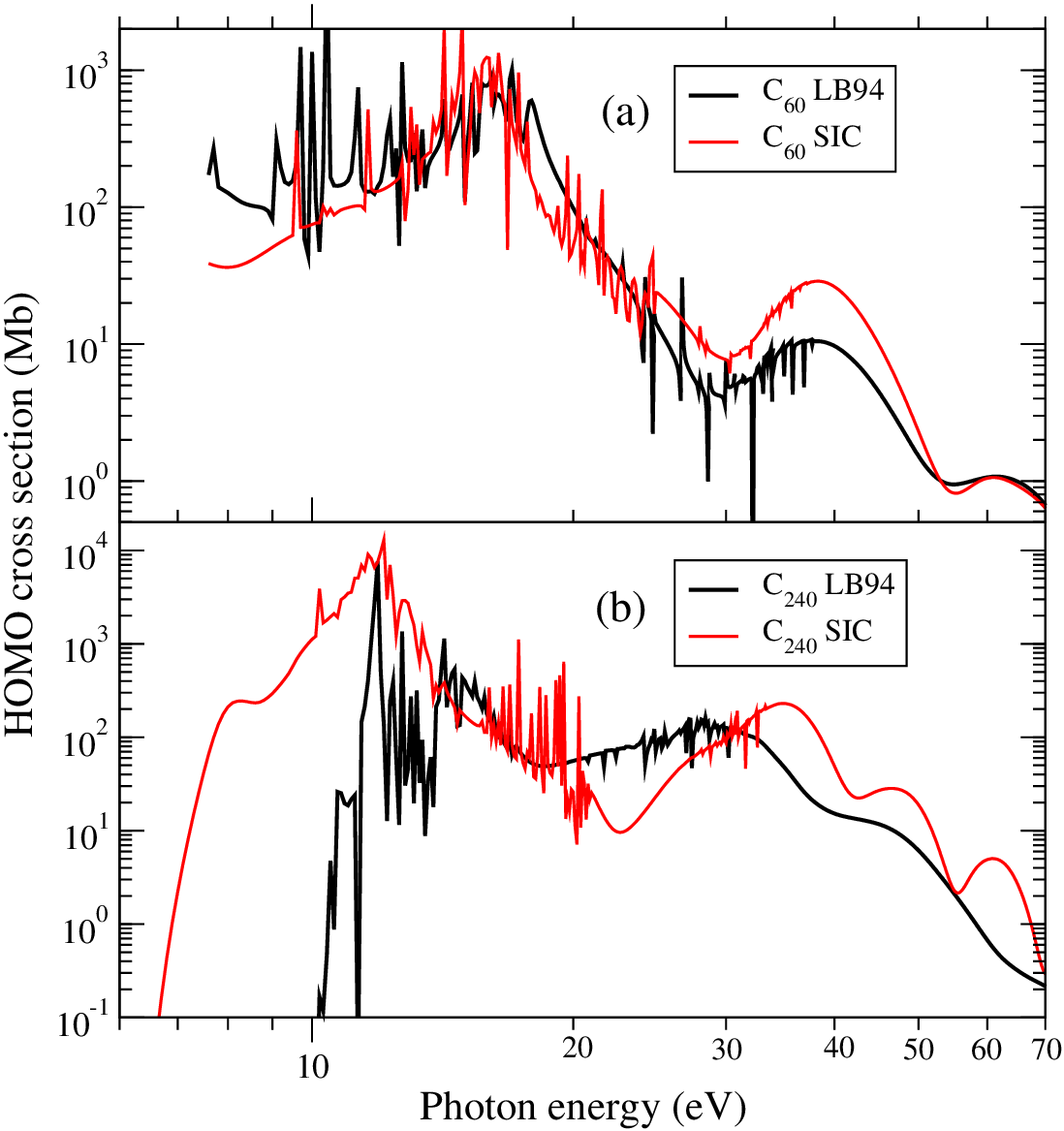}
\caption{(Color online) Photoionization cross sections for the HOMO level calculated in SIC and LB94 for C$_{60}$ (a) and C$_{240}$ (b).} \label{fig2}
\end{figure}
\begin{figure}[h!]
\includegraphics[width=8cm]{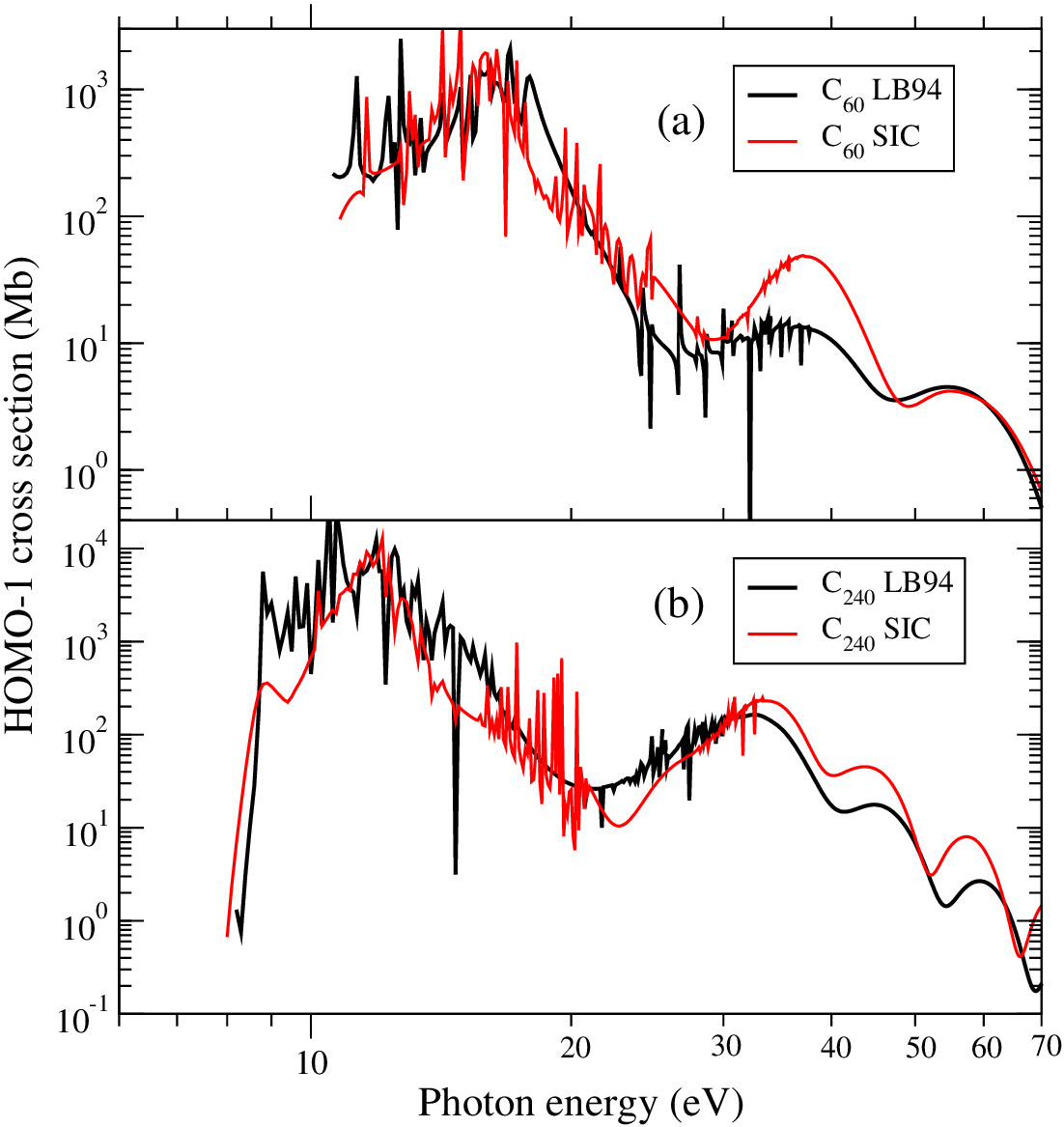}
\caption{(Color online) Photoionization cross sections for the HOMO-1 level calculated in SIC and LB94 for C$_{60}$ (a) and C$_{240}$ (b).} \label{fig3}
\end{figure}

\section{Results and Discussion}

\subsection{Photoionization of valence electrons}

The photoionization cross sections of the HOMO level calculated in TDLDA, both in LB94 and SIC schemes, are presented in Fig.\,2. Let us first note that the host of narrow spikes that appears represents single-electron autoionizing resonances. The positions and shapes of these resonances largely vary between two xc schemes This happens mainly because of their significantly different descriptions of the unoccupied excited states (that depend on the potential's asymptotic behavior), even though their occupied spectra are by and large similar. In fact, it is expected that owing to the better long-range accounts of electronic properties, LB94 resonances are more accurate in all current results. Neglecting these single-electron features, broad build-ups of the oscillator strength above 10 eV are due to the two collective plasmon resonances. The general shape of the curves is qualitatively similar between LB94 and SIC for C$_{60}$ [Fig.\,2(a)], largely because the HOMO levels are of the same $\pi$ symmetry in both the schemes [Fig.\,1(a)]. In contrast, due to the different symmetries of HOMO for C$_{240}$ [Fig.\,1(a)], the broad shapes of the LB94 and SIC curves in Fig.\,2(b) noticeably differ from each other. In all the curves, there appear some imposing oscillation-type structures above 30 eV that somewhat mask the second (40-eV) plasmon. Further, comparing Fig.1(a) with (b), we note a general shift of the plasmonic enhancements toward lower energies for larger C$_{240}$ similar to the known trend in the size dependence of plasmons in noble metal cluster studies~\cite{cottancin2006}.

TDLDA cross sections for HOMO-1 level are shown in Fig.\,3. Since for each fullerene the HOMO-1 level retains the same $\pi$ symmetry going from SIC to LB94, the broad shapes of the curves obtained from these approximations compare well, barring the mismatch in details including in the single-electron resonances. We also note here the superposed oscillatory structures at higher energies and the red-shift of the plasmon resonances in C$_{240}$ compared to C$_{60}$ as in the case of HOMO.

Cleaner shapes of the plasmon resonances are more readily captured in the total cross sections that we discuss in the next subsection. We address at this point a photoelectron diffraction-driven phenomenon that begins to surface from the waning region of the higher energy plasmon where the collective effect starts to weaken. An interference between photoelectron waves, predominantly produced at the boundaries of the fullerene shell, underpins this process. This essentially single-electron effect is the root cause of the oscillations seen at higher energies in Figs.\,2 and 3 that has been observed before in photoelectron spectroscopy~\cite{ruedel02,korica05} and theoretically discussed at great lengths~\cite{mccune2008}. Following Ref.\,[\onlinecite{mccune2008}], one can simply model these oscillations in a $n\ell$-level cross section by
\begin{eqnarray}\label{cross_osc}
\sigma_{n\ell\rightarrow k\ell'} &\sim& \frac{A^2(k)}{2}\left[B + (a_oh_o)^2\cos(2kR_o-\ell'\pi)\right. \nonumber \\
         &+& (a_ih_i)^2\cos(2kR_i-\ell'\pi) \nonumber \\
				 &-& 2a_oa_ih_oh_i\{\cos(2kR-\ell'\pi) + \left. \cos(k\Delta)\}\right],
\end{eqnarray}
where $A$ is a steady energy-dependent part, $a_o$ and $a_i$ are the values of the radial bound wavefunction at inner ($R_i$) and outer ($R_o$) radii of the fullerene shell, $B = a_o^2+a_i^2$, $h_i$ and $h_o$ are respectively proportional to the derivatives of the radial potential [Fig.\,1(b)] at $R_o$ and $R_i$, and $\Delta=R_o-R_i$. Obviously, the oscillations in photoelectron momentum ($k$) depend on the potential shape that also includes the angular momentum dependent centrifugal barrier. Therefore, it is not surprising that the higher energy sub-structures in Fig.\,2(a) and Fig.\,(3) qualitatively match between LB94 and SIC which have identical angular momentum symmetry. In Fig.\,2(b), however, this matching worsens. This is the consequence of increased centrifugal barrier from much higher angular momentum of LB94 HOMO level $1w$ for C$_{240}$ that obliterates the inner radius in the effective potential to effectuate $h_i = 0$ in \eq{cross_osc}, qualitatively altering the net shape of the oscillations. The details of this angular momentum effect were discussed earlier~\cite{mccune2008}. We must also note in Figs.\,(2) and (3) that these higher energy oscillations are in general smaller for C$_{240}$ as a consequence of the larger radius of this system leading to higher oscillation ``frequencies" in \eq{cross_osc}.  

Equation (\ref{cross_osc}) unravels some further insights. Note that the first three oscillatory terms in this equation carry a constant phase shift $\ell'\pi$, where the dipole selected final angular momentum $\ell'=\ell\pm1$. The implication is that each of these oscillations for ionization from two neighboring $\ell$ states will be 180$^o$ out-of-phase to each other~\cite{frank97}. However, the oscillation from $\sim a_oa_i\cos(k\Delta)$ in \eq{cross_osc} is independent of $\ell$. But note that between the ionization of a $\pi$ and a $\sigma$ electron this oscillation is roughly opposite, since the product $a_oa_i$ is negative for a $\pi$ radial wave, but positive for a $\sigma$ -- an implication of which will be discussed in the following subsection. 

\begin{figure}[h!]
\includegraphics[width=8cm]{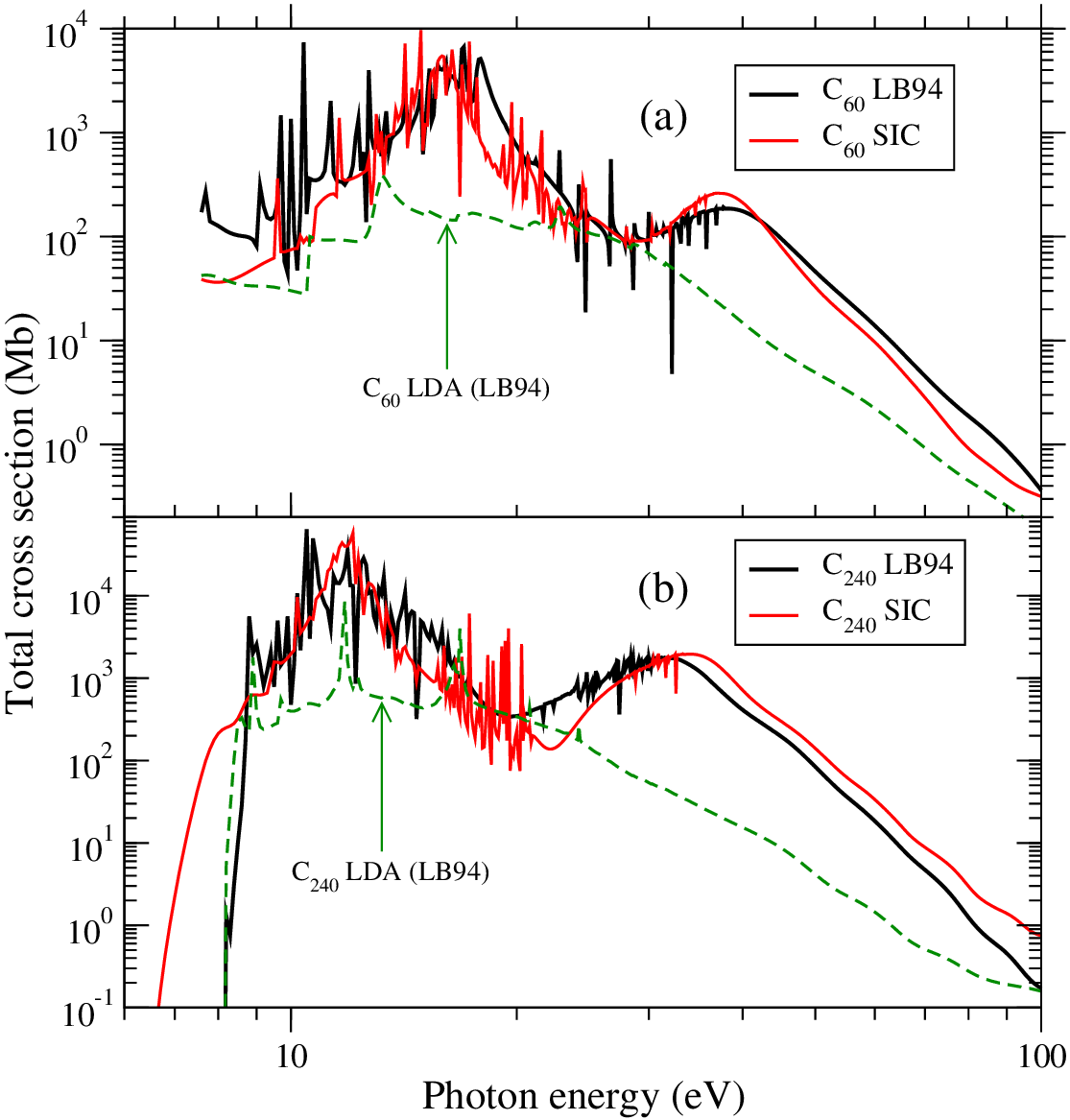}
\caption{(Color online) Total cross sections calculated in SIC and LB94 for C$_{60}$ (a) and C$_{240}$ (b). The corresponding single-electron (LDA) results using LB94 are also shown.} \label{fig4}
\end{figure}

\subsection{Total and band-differential cross sections}

Figure 4 presents the total TDLDA photoionization cross sections and compares them with respective single-electron LDA results (shown only for LB94) for both the fullerenes. The sum over $\ell$ largely cancels out oscillations due to the reason discussed above (in the last paragraph of subsection III A) and makes the broad higher energy plasmon (HEP) emerge clearly. In fact, both the plasmon resonances in TDLDA stand out in Fig.\,4 against the relatively smooth LDA curves. Unlike to the lower energy plasmon (LEP) resonances, HEPs exhibit far weaker effects of single-electron resonances but rather long decay tails. Energy red-shifts of the resonances in C$_{240}$ compared to those in C$_{60}$ are noted along with the fact that C$_{240}$ plasmons are utilizing significantly higher oscillator strength density due to its much larger electron pool to collectivize. For each fullerene, significant differences in the resonance widths between LB94 and SIC are also noted. Values of various resonance parameters are given in Table II.

From a non-perturbative many body theory, the emergence of plasmon resonances can be thought of as originating from the formation of collective excited states under the influence of external electromagnetic field~\cite{zangwill80}. Since the collective excitations are energetically embedded in the single-electron ionization channels, they provide alternative ionization pathways degenerate with single-electron channels.  Thus, the {\em autoionization} of these collective excited states induce resonant enhancements in the subshell cross sections as shown in Figs.\,2 and 3. However, from a perturbative approach the plasmon mechanism can be best modeled by Fano's interchannel coupling formalism~\cite{fano61}. To include the effects of channel-coupling upon the final state wave function of each of the perturbed dipole matrix elements $M_{n\ell}(E)$ one can write~\cite{madjet-jpb2008},       
\begin{widetext}
\begin{equation}\label{mat_element}
M_{n\ell}(E) = D_{n\ell}(E) +  \displaystyle\sum_{n'\ell' \neq n\ell} \int dE' \frac{\langle\psi_{n'\ell'}(E')|\frac{1}{|{\bf r}_{n\ell}-{\bf r}_{n'\ell'}|}
|\psi_{n\ell}(E)\rangle}{E-E'} D_{n'\ell'} (E')
\end{equation}
\end{widetext}
where $D_{n\ell}$ is the unperturbed (LDA) $n\ell$ matrix element, $\psi_{n\ell}(E)$'s are the unperturbed final continuum channel wave functions of the single-electron channels, and the sum is over all of the photoionization channels except the $n\ell$ channel. The matrix element within the integral of Eq.\,(\ref{mat_element}) is known as the interchannel coupling matrix element; the fact that each of $n\ell$ initial state orbital overlaps strongly with all other fullerene-orbitals insures that these interchannel coupling matrix elements will be strong.  Further, this also justifies the existence of both low and high energy plasmons at exactly the same energies for all the subshells for a given fullerene and implies the various dipole matrix elements are ``in phase'' over the two energy regions (bands) of each fullerene [Fig.\,1(a)]. Consequently, the various terms in the sum in Eq.\,(\ref{mat_element}) will add up {\em coherently}, leading to the dramatic enhancement. 
\begin{figure}[h!]
\includegraphics[width=8cm]{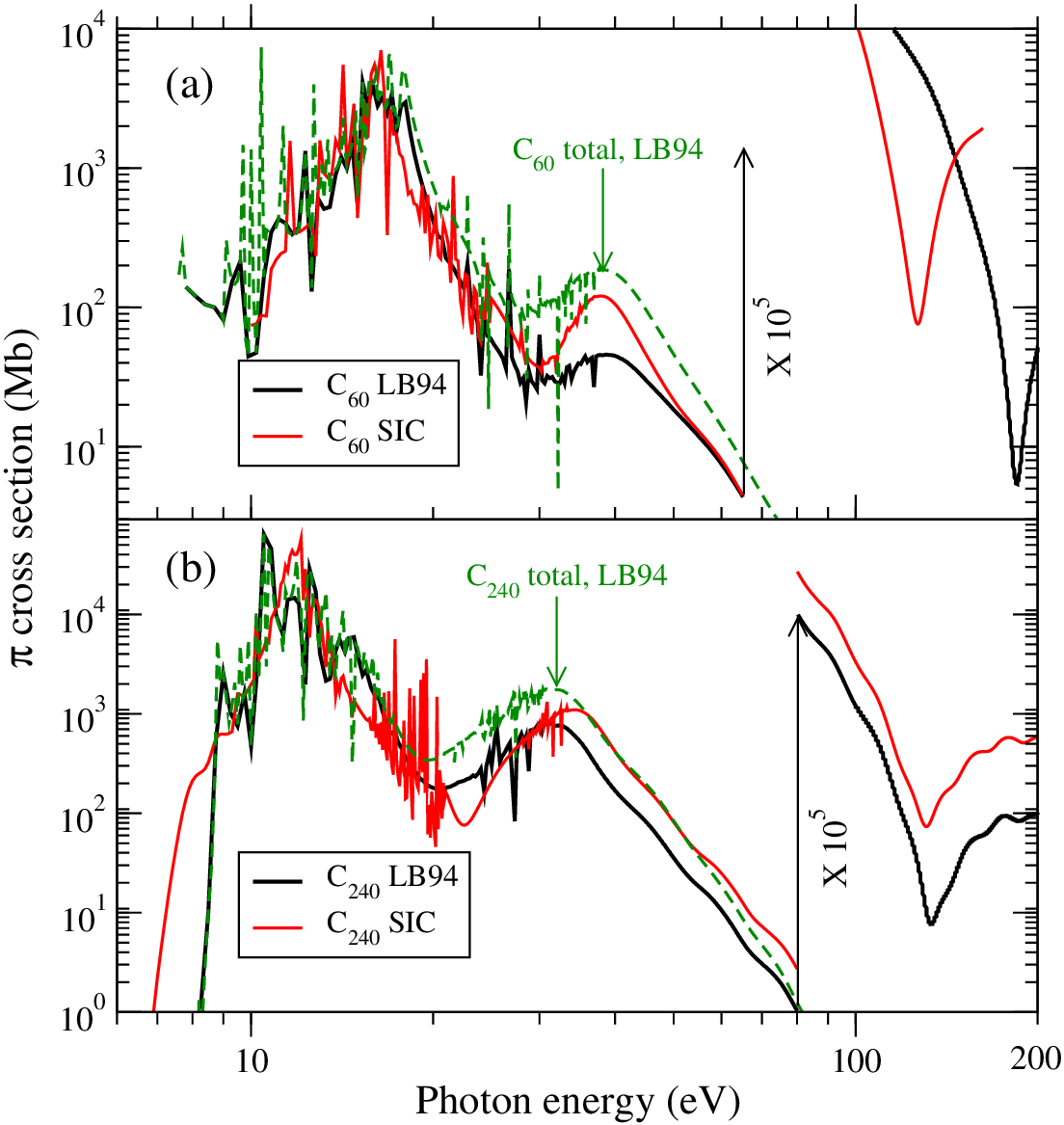}
\caption{(Color online) Total $\pi$-band cross sections in SIC and LB94 for C$_{60}$ (a) and C$_{240}$ (b). Curves are scaled at higher energies to illustrate a strong minimum (see text). The total cross sections are also displayed for comparisons.} \label{fig5}
\end{figure}
\begin{figure}[h!]
\vspace{1 cm}
\includegraphics[width=8cm]{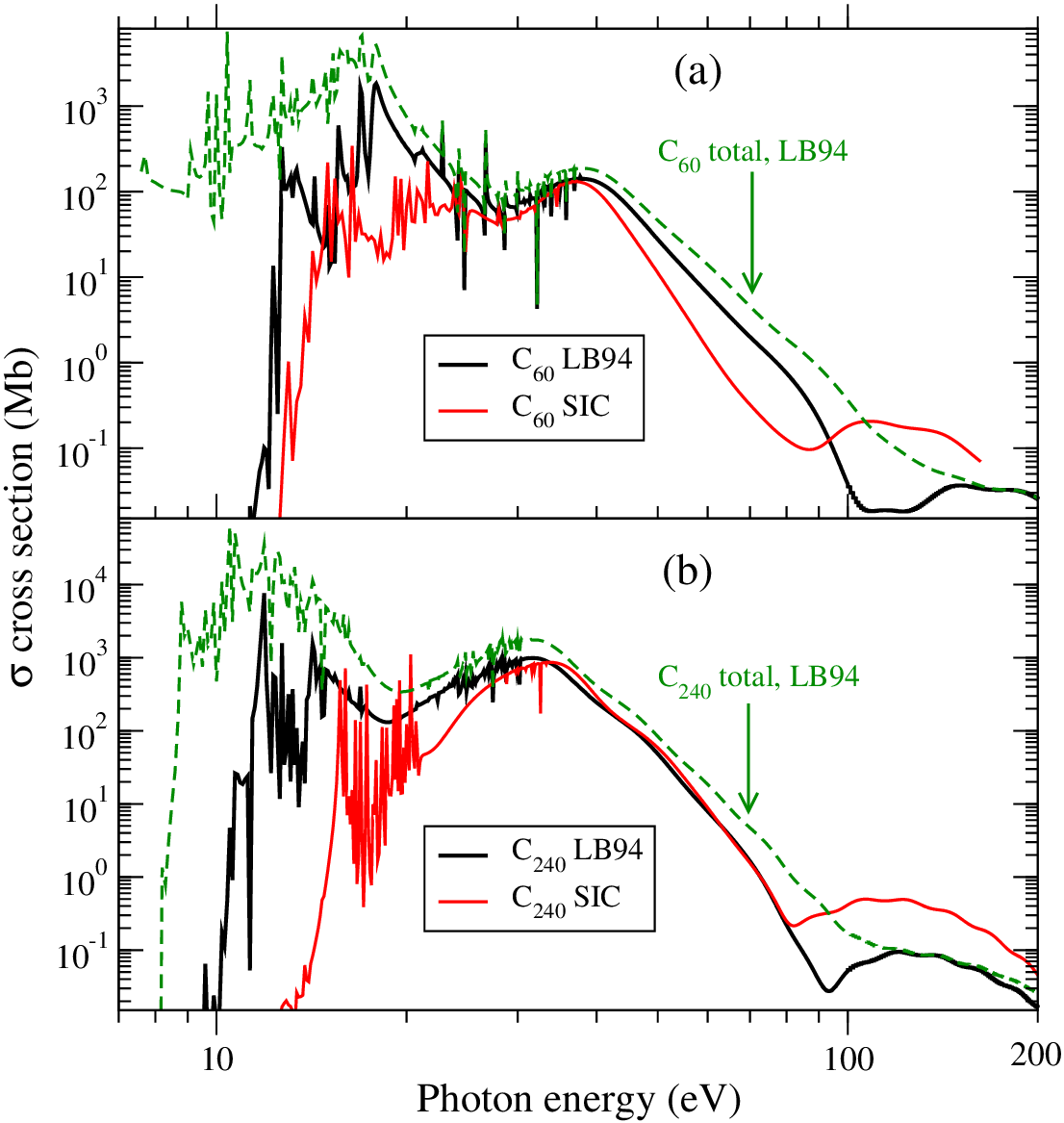}
\caption{(Color online) Total $\sigma$-band cross sections in SIC and LB94 for C$_{60}$ (a) and C$_{240}$ (b). The total cross sections are also displayed for comparisons.} \label{fig6}
\end{figure}

Equation (\ref{mat_element}) reveals one further important correlation feature. Since a $\pi$ (or a $\sigma$) bound orbital will have near-perfect overlaps with other $\pi$ (or $\sigma$) orbitals due to their almost identical shape and spatial extent, the interchannel coupling matrix element in \eq{mat_element} will be stronger for a $\pi$-$\pi$ or a $\sigma$-$\sigma$ self-coupling than a $\pi$-$\sigma$ cross-coupling. Therefore, it is expected that the $\pi$ electrons will show a preferred participation for building LEP and the $\sigma$ for HEP. Figs.\,5 and 6 respectively show the only-$\pi$ and only-$\sigma$ band cross sections in TDLDA calculated in both LB94 and SIC. For each fullerene, if we compare the $\pi$ band result in LB94 with the total cross section in LB94 (also shown), a dominant contribution of the $\pi$-cloud to LEP and of the $\sigma$-cloud to HEP are indeed noted. In general, however, it is also obvious from these comparisons that either of LEP and HEP in a fullerene are of both $\pi$ and $\sigma$ mixed character, it is just that one is dominant on the other.

A discussion on the red-shift of TDLDA plasmon resonances from C$_{60}$ to C$_{240}$ [Fig.\,4] may now be in order. Classical plasmon-model~\cite{lambin92} of a spherical dielectric shell with symmetric and antisymmetric vibrations between the inner and outer surfaces suggests that the midpoint energy between the two resonances to be about the same for C$_{60}$ and C$_{240}$, since they have approximately same initial electron densities~\cite{korol07}. According to this model the plasmons are then formed below and above this midpoint energy shifted equally both ways, and this shift grows with the increasing radius, suggesting that the plasmons will be more separated out for C$_{240}$~\cite{korol07}. Clearly, that is not seen in Fig.\,4, in which both the plasmons red-shift for C$_{240}$ and in fact move close to each other compared to their C$_{60}$ results (see Table. II for the actual values), suggesting that quantum effects play an important role. One possible way to understand this phenomenon quantum mechanically is to recall in Fig.\,1(b) that the C$_{240}$ ground state potential is shallower while accommodating a number of electrons four times that of C$_{60}$ producing far compact energy levels. This suggests a decrease of the {\em average} ground state binding energy for C$_{240}$. Therefore, since in the spirit of \eq{mat_element} the plasmons can be interpreted as the coherence in close-packed single-electron excitations, it is only expected that the plasmons will begin to excite at lower photon energies causing their early onsets for C$_{240}$, as seen in Fig.\,4. In fact, this trend of red-shifting plasmons with increasing fullerene size should be rather generic, at least in the jellium based quantum calculations. An insight in the phenomenon can be motivated by perceiving a collective mode as having a natural oscillation frequency $\sqrt{\kappa/\rho}$ of a mass density ($\rho$) on a spring of stiffness $\kappa$~\cite{mccune2011}. Thus, a shallower binding potential with higher electron population for C$_{240}$ translates to the loosening of the spring decreasing $\kappa$ and thereby its resonant frequencies.

Let us now compare between the predictions of LB94 and SIC for the band-cross sections. For the $\pi$-band, LB94 retains a contribution approximately similar to that of SIC at LEP, but shows depletion at HEP which is more prominent for C$_{60}$ [Fig.\,5(a)] than C$_{240}$ [Fig.\,5(b)]. For the $\sigma$-band, on the other hand, a notably higher contribution to LEP and some increase at HEP by LB94 for both the systems are found [Fig.\,6]. There are more. Our discussions following \eq{cross_osc} indicate that the $\ell$-sum over $\pi$ or $\sigma$ cross sections will significantly weaken the diffraction oscillations coming from the first three oscillatory terms in \eq{cross_osc}, while the fourth oscillation $\sim\cos(k\Delta)$ will survive being free of $\ell$. As a result, in the band-cross sections this $\Delta$-dependent oscillation will dominate. Since $\Delta$ slightly shortened in LB94 than SIC for C$_{60}$ [Fig.\,1(b)], $\pi$-band LB94 curve in Fig.\,5(a) produces a longer wavelength in $k$ to induce its first minimum above 100 eV at an energy higher than that in SIC. The equality of $\Delta$ in LB94 versus SIC for C$_{240}$, on the other hand, justifies the occurrence of these minima at about the same energy as in $\pi$-band results for this fullerene [Fig.\,5(b)]. However, this effect is not so intuitive for the $\sigma$-band case. As seen, while the minimum in LB94 for C$_{60}$ [Fig.\,6(a)] does appear at higher energies than the SIC minimum, they do not seem to coincide for C$_{240}$ [Fig.\,6(b)] as they did for C$_{60}$. The latter is due to the fact that the $\sigma$ states for C$_{240}$, reaching very high in $\ell$ values compared to their counterparts in C$_{60}$, produce such strong centrifugal repulsions that the effective potentials for high $\ell$ considerably deform rendering the role of $\Delta$ less meaningful~\cite{mccune2008}.
\begin{figure}[h!]
\vspace{1 cm}
\includegraphics[width=8cm]{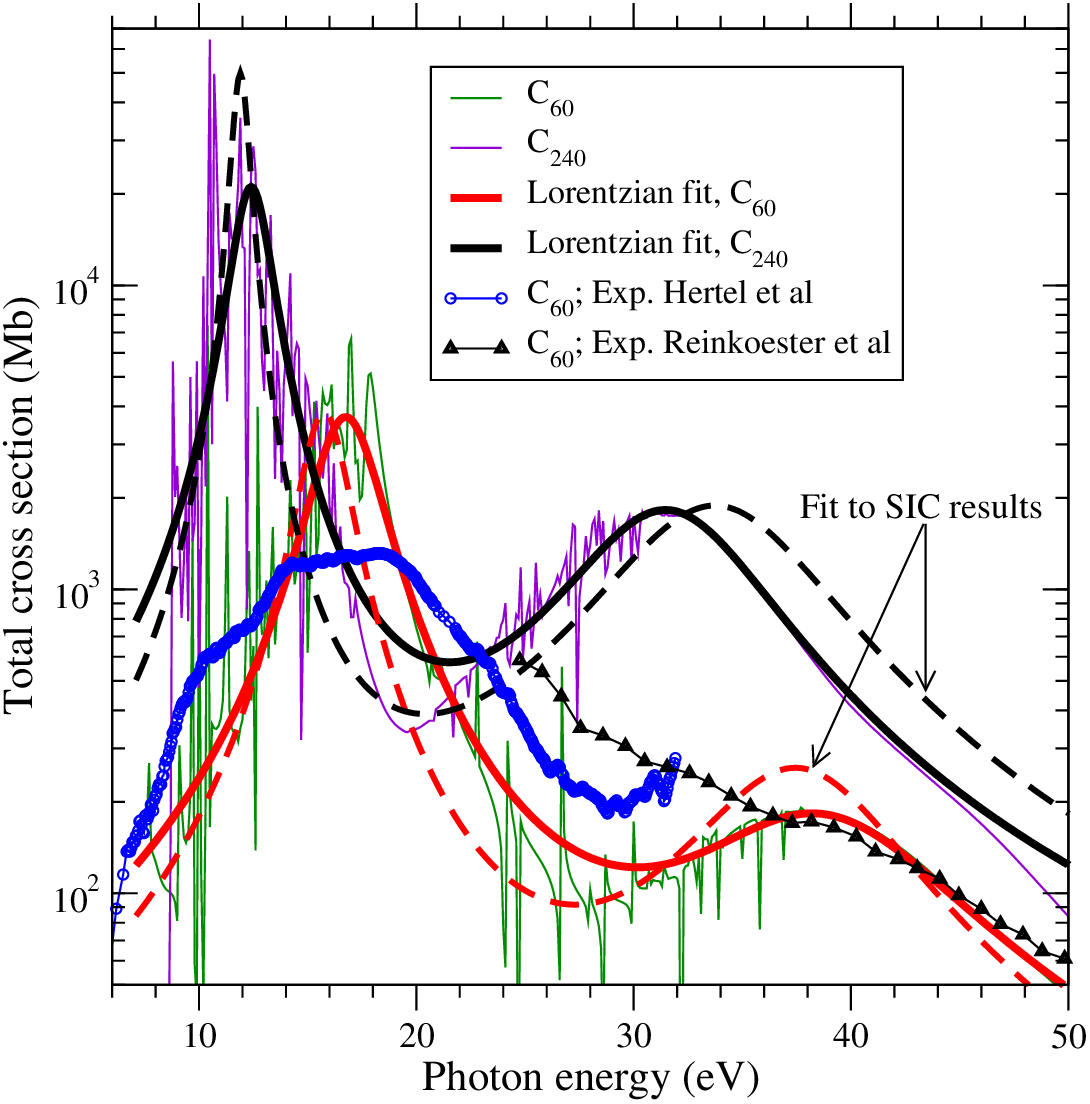}
\caption{(Color online) Lorentzian fits to the total cross sections (also shown) obtained from LB94. These are compared with the fits of the corresponding SIC results. Two sets of experimental data, appropriately red-shifted, are included to aid the comparison between plasmonic responses obtain via two xc schemes.} \label{fig7}
\end{figure}

\subsection{Plasmon resonances and comparison with experiments}

Measurements~\cite{scully05,scully07,rein04} of plasmon resonances in the photoionization of neutral and ionic C$_{60}$ produced relatively smooth curves without any evidence of autoionizing resonances, which exist in our theoretical TDLDA results. As discussed in details in our earlier study~\cite{madjet-jpb2008}, this is likely because the coupling of electronic motions with the temperature-induced vibration modes of the core~\cite{bertsch89} and the fluctuation of the cluster shape around the shape at absolute zero~\cite{penzar90,pacheco89}. In addition, the inherent over-delocalization of jellium models predicts autoionizing resonances that are too narrow, as seen in our results. Therefore, in Fig.\,7, we fit the non-spiky background parts of our TDLDA total cross sections obtained via LB94 for both the fullerenes using a formula that includes two Lorentzian line profiles. We further present in Fig.\,7, two similar fitting curves for the SIC results of the fullerenes. For both fullerene systems, Table 2 presents the positions ($E_o$) and full-widths at half-maxima (FWHM), $\Gamma$, and oscillator strength densities (OSD) corresponding to each plasmon resonances calculated in LB94 and SIC; LB94 results are displayed in parenthesis.
%
\begin{table}
\caption{Resonance positions ($E_o$), FWHM ($\Gamma$) and oscillator strength density (OSD) of the lower energy (LEP) and the higher energy (HEP) plasmon. The values in parenthesis are corresponding LB94 results.}
\label{tab2}       
\begin{tabular}{c|ccc}
\hline\noalign{\smallskip}
              & $E_o$ (eV)   & $\Gamma$ (eV) &  OSD \\
\noalign{\smallskip}\hline\noalign{\smallskip}
C$_{60}$ LEP  & 15.8 (16.8)  & 2.5 (3.5)     & 136 (184) \\
C$_{60}$ HEP  & 37.5 (38.5)  & 10.0 (13.0)   & 35 (30)   \\
\noalign{\smallskip}\hline\noalign{\smallskip}
C$_{240}$ LEP & 11.9 (12.4)  & 0.9 (2.0)     & 642 (601) \\
C$_{240}$ HEP & 33.8 (31.5)  & 10.5 (9.5)    & 281 (241) \\   
\noalign{\smallskip}\hline
\end{tabular}
\end{table}

From Table 2 and Fig.\,7, going from SIC to LB94, both LEP and HEP of C$_{60}$ move up in energy by 1 eV, while C$_{240}$ LEP by a half eV. We recall the spirit of a classical oscillator model of dielectric shell that the plasmon frequencies are proportional to the square-root of the ratio of rigidity to density ($\sqrt{\sim \kappa/\rho}$, in analogy to the oscillation frequency of a mass on a spring of stiffness constant $\kappa$) introduced in the previous subsection. Note, LB94 radial waves being slightly more spread out than their SIC counterparts occupy a larger space effectively decreasing the density. This explains the blue-shift of LB94 plasmons. This trend in LB94 is an improvement, since jellium based prediction of C$_{60}$ plasmon resonance energies are known to be below their measured values~\cite{madjet-jpb2008}. However, this trend is reversed for C$_{240}$ HEP where LB94 moves this plasmon lower in energy by more than 2 eV from its SIC prediction, ascertaining the importance of quantum effects to capture the details of these resonances. Furthermore, the LB94 width of C$_{60}$ LEP is found to be 3.5 eV, which is an increase of 40\% over its SIC value of 2.5 eV, while this increase is 30\% for C$_{60}$ HEP. More than a double increase of width for C$_{240}$ LEP is found going from SIC to LB94, while again, this trend reverses by a small amount for C$_{240}$ HEP. Significant variations in the OSD utilized by each plasmons for either system between two xc approximations are also noted in Table 2, accounting for the detailed differences that the two calculation schemes generate.

Comparisons of the results between the two fullerenes in Fig.\,7 as well as in Table 1 indicate a generic red-shift of plasmon energies for the larger fullerene C$_{240}$, as noted and discussed earlier. We also find in Table 2 a general trend of the width $\Gamma$ to decrease with the increasing size of fullerene, except for C$_{240}$ HEP in SIC. Further note that while for C$_{60}$ LEP the OSD value increases from SIC to LB94, the trend is found opposite for this resonance of C$_{240}$. For the HEP, either fullerene exhibits decrease in OSD going from SIC to LB94.

Fig.\,7 further includes two sets of experimental measurements for C$_{60}$, where the data from Hertel et al~\cite{hertel92} are red-shifted by 3 eV and those from Reink\"{o}ster et al~\cite{rein04} by 1 eV to match respectively with the energies of LEP and HEP calculated in LB94. As evident, the modifications in $\Gamma$ and OSD, as brought about by the LB94 scheme, indicate an improved agreement with experimental results compared to what SIC achieves. We must also note that in a jellium model, the plasmon resonances only decay via the degenerate single-electron channels. In the real system, however, there would be additional effects from the independent local ion sites positioned based on an appropriate atomistic symmetry, at least for relatively more tightly bound electrons. As shown in detailed with SIC results in Ref.\,[\onlinecite{madjet-jpb2008}], in order to account for these additional decay channels, the theoretical cross section in a jellium frame must be convoluted with a small width in order for a more meaningful comparison with measurements. With the already improved agreement of the current ``zero-width" results of LB94, it is only expected that such a convolution will further better the agreement with the experiment. 
  
\section{Conclusions}      

In conclusion, the work accounts for various robust similarities but detailed differences between the results obtained via two standard xc schemes, SIC and LB94, in the framework of density functional description of delocalized valence electrons of the fullerene molecule where the ionic core is treated as a jellium shell. The focus has been applied to understand both the ground state and single-photoionization properties of the system. For the ionization study, the ultraviolet energy range of plasmon activities and above-plasmon soft x-ray range were considered. The comparison between the results of two prototypical spherical fullerenes, C$_{60}$ and C$_{240}$, further unravels the scopes of validity of these two theoretical schemes. A natural next step is to consider the influence of xc formalism on the photospectroscopy of non-spherical fullerenes, which, however, is a topic for our future research. To this end, within the known limitation of the jellium description of the molecular ion-core, the gradient corrected LB94 formalism seems to bring the results closer to the measurements on C$_{60}$ over the plasmon resonance energy region. We hope that with possible future experiments with C$_{240}$ the success of LB94 scheme can be verified for larger fullerene systems as well. 

\begin{acknowledgments}
The work is supported by the National Science Foundation, USA.
\end{acknowledgments}

\end{document}